\documentclass[preprint]{aastex}
\newcommand{\microns}{\micron}
\begin{document}

\title{A Mid-Infrared Study of the Class 0 Cluster in LDN 1448}

\author{JoAnn C. O'Linger\altaffilmark{1}}
\affil{Spitzer Science Center, California Institute of Technology,
MS 314--6, Pasadena, CA 91125}
\email{joanno@ipac.caltech.edu}
\altaffiltext{1}{Visiting Astronomer at the Infrared Telescope Facility,
which is operated by the University of Hawaii under contract from the
National Aeronautics and Space Administration.} 

\author{David M. Cole\altaffilmark{1}}
\affil{Jet Propulsion Laboratory, 4800 Oak Grove Dr., 
MS 306--388, Pasadena, CA 91109}
\email{David.M.Cole@jpl.nasa.gov}

\author{Michael E. Ressler}
\affil{Jet Propulsion Laboratory, 4800 Oak Grove Dr., 
MS 79-5, Pasadena, CA 91109}
\email{Michael.E.Ressler@jpl.nasa.gov}

\and

\author{Grace Wolf-Chase\altaffilmark{2}}
\affil{Dept.~of Astronomy \& Astrophysics, University of Chicago,
5640 S.~Ellis Ave., Chicago, IL 60637}
\email{grace@horta.uchicago.edu}
\altaffiltext{2}{Joint appointment at Adler Planetarium and Astronomy Museum,
1300 S. Lake Shore Drive, Chicago, IL 60605}

\begin{abstract}

We present ground-based mid-infrared observations of Class~0 protostars
in LDN 1448. Of the five known protostars in this cloud, we detected two,
L1448N:A and L1448C, at 12.5, 17.9, 20.8, and 24.5~\microns, and a
third, L1448 IRS~2, at 24.5~\microns.  We present high-resolution
images of the detected sources, and photometry or
upper limits for all five Class~0 sources in this cloud.
With these data, we are able to augment existing spectral
energy distributions (SEDs) for all five objects
and place them on an evolutionary status diagram.
\end{abstract}


\section{Introduction}\label{sec:intro}

The majority of young stellar objects (YSOs) are classified using a scheme based on 
characteristics of their SEDs introduced by \citet{lad84} 
\citep[see also][]{lad87,lad91}.  They proposed three classes: I, II, and III; distinguished 
from one another by their near-infrared to mid-infrared spectral indices, which 
were presumed to be indicators of the evolutionary status of the object in question.
Class I sources were initially thought to be representative of the earliest
stage of pre-main-sequence evolution, {\em i.e.} protostars.

The spectral index, $\alpha$, is defined as

\begin{equation}
\alpha = -\frac{\mbox{d}\,\log{\nu F_{\nu}}}{\mbox{d}\,\log{\nu}};
\end{equation}

\noindent for purposes of YSO classification, the region of the SED usually used for
this calculation is from 2.2 $\mu$m to 10 $\mu$m. The spectral index is 
the negative of the slope of the SED, when graphed in the frequency domain \citep{lad91}. 

The most current version of this slope classification scheme is presented in 
\citet{gre94}. For a Class 
I source, the SED tends to be broader than a single-temperature blackbody
and $\alpha > 0.3$, due to the fact that the peak of the SED of such
a source is shifted towards the far-infrared because of the very low effective 
temperature of the YSO.
These sources are usually deeply embedded in their nascent dust envelopes.
For Class II and Class III YSOs, on the other hand, 
 $ -0.3 > \alpha \ge -1.6$ and $\alpha < -1.6$, respectively.
The SED of a Class II will be much broader than that of a Class III, due to 
the presence of circumstellar dust with a fairly wide temperature distribution.  
\citet{gre94} identify a fourth class of ``flat spectrum'' sources,
for which $0.3 > \alpha \ge -0.3$, that have near-infrared spectra which are
strongly veiled by continuum emission from hot circumstellar dust.

Observations of candidate protostars, in particular VLA 1623 in $\rho$
Ophiuchi A, an extremely cold ($T_{e\!f\!\!f} \leq$ 20 K) object that does
not appear at {\em IRAS} wavelengths, led to the introduction of a new
class of young stellar objects (YSOs) by \citet{an93}.  They contended that
Class I sources could not be the youngest examples of pre-main-sequence 
objects ({\em i.e.}, not true protostars), and promptly created a new category 
for such objects, designated Class 0.  The search
for these protostars has gained impetus in the last several years due to
the development of highly sensitive submillimeter continuum detectors. 

Class~0 protostars are the rarest of YSOs. Pre-main-sequence stars typically spend 
$\ll 10^5$ years in this phase, on their way towards the onset of hydrogen fusion 
and stellar birth.  The current set of defining characteristics of a Class 0 
protostar are as follows \citep{an00}:

\begin{itemize} 
\item Detection of a radio continuum source, presence of a molecular outflow, or 
some other evidence of a central protostellar source; 
\item Centrally peaked but extended submillimeter continuum emission indicating an 
envelope as opposed to a mere disk;
\item A ratio of L$_{submm}$/L$_{bol}$ $>$ 0.5\% (where L$_{submm}$ is the luminosity 
radiated by the object at wavelengths longward of 350 $\mu$m and L$_{bol}$ is the total 
luminosity of the object).
\end{itemize} 

This last criterion indicates a circumstellar mass greater than the mass of the central 
protostellar core, as discussed in \citet{an93}. There is no mention of the spectral index
between 2 and 10 $\mu$m, since these sources have generally not been detected shortward of 
10 $\mu$m using ground-based instruments. This will change, no doubt, as detector technology 
evolves.

Although reports of newly discovered Class~0
protostars have increased dramatically in the last few years,
confirmed sources in this category still number well under a hundred.
\citep{an00,fro05}. The spectral energy distribution (SED) of a Class~0 object,
with a typical effective temperature of $\leq$40~K, peaks at around
100~\microns; until fairly recently, essentially the only tool with which one may study
the Wien side of the SEDs of protostars has been the \emph{IRAS} database.

\begin{figure}[t]
\centering
\includegraphics[width=18.0cm]{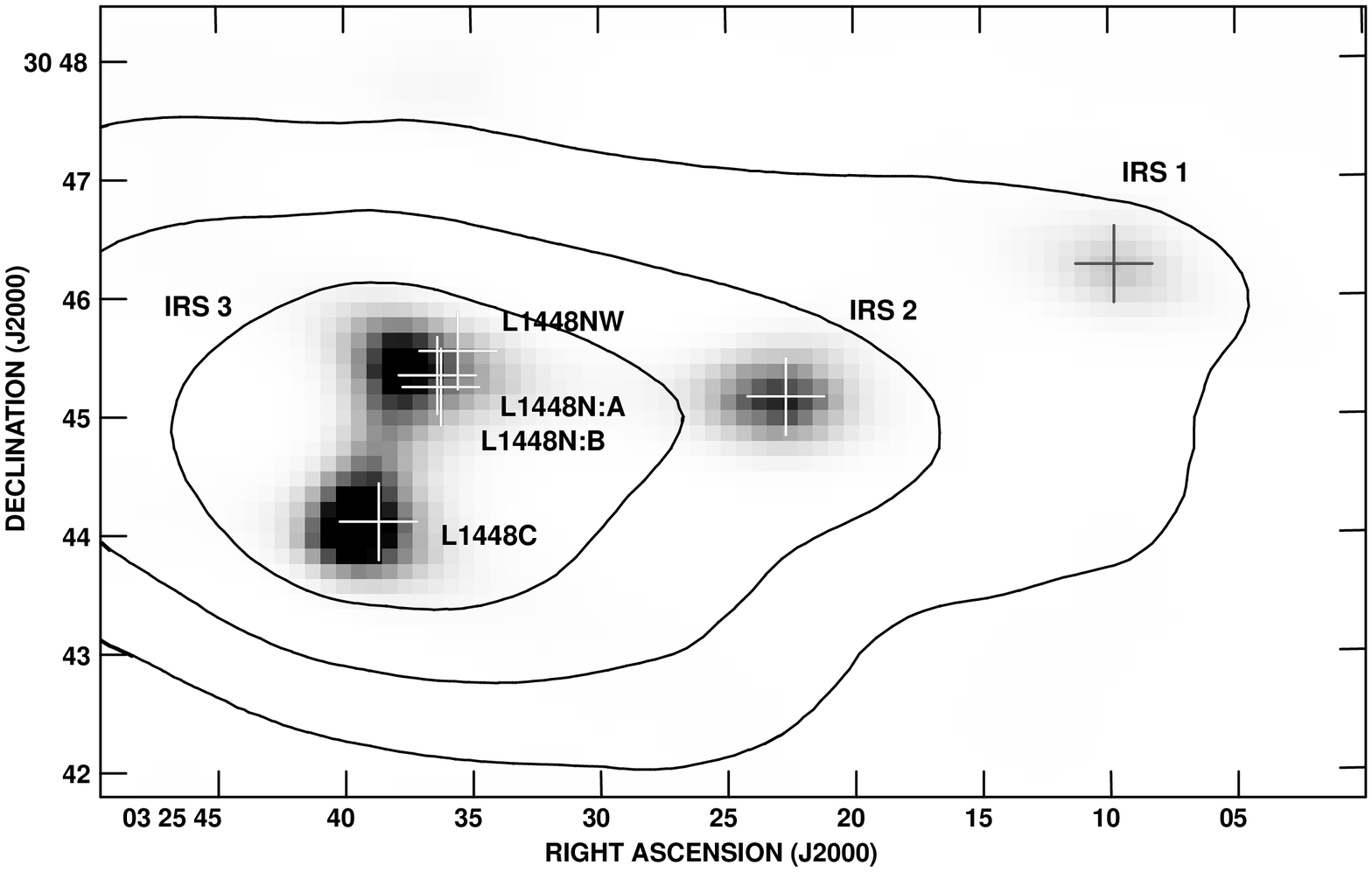}
\caption{{\bf \emph{IRAS} map of L1448:} 
An {\em IRAS} image shows the three Point Source Catalog objects in L1448,
also known as L1448 IRS~1, IRS~2, and IRS~3. Contours represent the
unenhanced coadded \emph{IRAS} data, overlaid on the greyscale plot of the
corresponding HIRES data. Crosses indicate the known positions of
young stellar objects from high-resolution observations at other
wavelengths. The crosses are labeled with the most commonly used
nomenclature for these sources found in the literature.\label{fig:hires}}
\end{figure}

The majority of known Class~0 sources were detected by \emph{IRAS} at
25~\microns, if not at 12~\microns; however, due to the nature of the
\emph{IRAS} data, the effective beam size at mid-infrared wavelengths
is never less than $\sim$30--40\arcsec,
even when using the high-resolution data-processing program YORIC
\citep{aum90,hur96,bar98,oli99}. Figure
\ref{fig:hires} provides an example of coadded 60-\micron\ \emph{IRAS} data
before (contours) and after HIRES processing (greyscale) via YORIC. The
poor spatial resolution of the \emph{IRAS} data makes associating point
sources in the catalog with data from other wavelengths difficult and, in
cases of close binary or multiple systems, does not allow the contributions
of the individual components to the total flux to be distinguished. HIRES
point-source modeling, which makes use of a little-known feature of the YORIC
program, has been used successfully to estimate individual \emph{IRAS}
fluxes from sources as close as 20\arcsec\ apart, but this still does not
compare with the sub-arcsecond resolution obtainable with millimeter-wave
interferometry \citep{bar98}. Consequently, the mid-infrared spectral
energy distributions of Class 0 sources have tended to be poorly constrained.
Data from the latest NASA Great Observatory, the \emph{Spitzer Space Telescope}
are already contributing significantly to our understanding of the mid-IR
characteristics of young stellar objects, but \emph{Spitzer} lacks the resolution to 
augment the information on the small-scale structure of their circumstellar
environments which may be obtained from submillimeter and
millimeter wavelength observations \citep{ch00, lo00}.

Some of the most important questions in star formation studies
today center around the multiplicity of these young systems, and how the
binary properties evolve with time \citep{lo00,rei00}. The majority of
nearby young T Tauri stars are known to be binary systems, yet only a very
few Class~0 protobinary systems have been resolved. It has been postulated
that the giant Herbig-Haro flows driven by many young stellar objects may
be initiated by the dynamical decay of unstable multiple systems
\citep{rei00}. Understanding star formation in the context of multiple systems has
important implications for understanding the IMF and its physical meaning.

In order to address the issue of multiplicity and obtain details about the
immediate circumstellar environments of Class~0 sources, it is necessary to
study these objects at as many wavelengths as possible with high angular 
resolution. 

A particularly rich cluster of Class~0 sources is located in the compact globule
L1448, a molecular cloud which is part of the Perseus molecular complex.
L1448 lies at a distance of $\sim$300~pc; its mass, estimated from ammonia data, 
is $\sim$50~M$_{\sun}$ \citep{ba86a}.
Three embedded infrared sources were detected by \emph{IRAS} in this cloud;
since then, it has been surveyed at several wavelengths and mapped in a 
number of molecular lines \citep{ba86b,ang89,bal97,ei00,wo00a}.
Listed from west to east, the \emph{IRAS} Point Source Catalog designations 
for the three sources are IRAS 03220+3035, IRAS 03222+3034, and IRAS 03225+3034. 
In the literature, these objects are more commonly known as L1448 IRS~1, L1448
IRS~2, and L1448 IRS~3, respectively (see Figure \ref{fig:hires}).

Only the western-most of these objects (IRS~1) has been detected at
near-infrared and shorter wavelengths \citep{co79,ne84}; this source has
been identified as a Class I YSO and hence was not included in our survey 
\citep{ei00,bar00}.
The other two \emph{IRAS} point sources have been found to contain multiple
Class~0 objects. IRS~2 was recently reported by \citet{wo00a} to be a candidate 
protobinary system, based on the two distinct molecular outflows which emerge 
from that position, although the binary components were unresolved at that time. 
Subsequently \citet{vo02} announced that both components of the
IRS~2 binary had in fact been detected at BIMA. 
IRS~3 consists of L1448N:A,B (a 7\arcsec\ protobinary), 
L1448NW, and L1448C (note: L1448C is frequently referred to as L1448-mm), 
all of which drive molecular outflows as well
\citep{ter97,bar98,ei00,wo00a}. The locations of all of these sources, from
previously published high-resolution submillimeter or millimeter-wave
observations, are indicated by crosses in Figure \ref{fig:hires}
\citep[and references therein]{oli99,lo00,an00}.

We observed all of the known L1448 Class~0 protostars in the mid-infrared, and 
present herein images of the detected sources plus multi-wavelength photometry 
and/or upper limits for all five Class~0 sources.

\section{Observations and Data Reduction}\label{sec:obs}

Low-mass Class~0 objects are generally not visible at wavelengths shorter
than 10~\microns\ using current (ground-based) technology \citep{an93,bar94}.
The shortest wavelengths at which these sources are generally thought
to be easily detectable are in the $\lambda$ $\sim$ 20~\micron\ 
atmospheric window. MIRLIN, JPL's 128x128~pixel Si:As mid-infrared camera 
\citep{ress94}, provides access to this wavelength regime.

The observations reported here were carried out using MIRLIN at NASA's
3-m Infrared Telescope Facility (IRTF), on the nights of 2000 Feb 18--19 UT
and 2000 Oct 5 UT. For these observations, the measured pixel scale was
$0\farcs482$, yielding a field of view just over 1\arcmin\ on a side. To
remove the thermal background signal from the sky and the telescope, the
secondary mirror was chopped 28\arcsec\ SE--NW at about a 5 Hz rate, and
the telescope nodded the same distance and direction every few hundred chop
cycles. The L1448C source and the L1448N/NW system were both observed at
12.5, 17.9, 20.8, and 24.5~\microns, while L1448 IRS~2 was observed using
only the latter two filters. $\alpha$~Aur,
$\alpha$~Boo, and $\alpha$~CMi were observed
throughout each night in February, in all four passbands, both to serve as
photometric calibrators and to track the weather conditions and airmass
dependence. In October, $\alpha$~Lyr was used for the same
purposes. The weather was excellent in February but more variable in
October.

The data were reduced using our in-house IDL routine, MAC (Match And
Combine), which performs the background subtraction, identification and
shifting of the source peaks, and the coaddition of the registered data.
Circular aperture photometry was then performed on both the standards and
the L1448 sources using apertures of radius 5 pixels ($2\farcs4$).

\section{Results}\label{sec:res}

\begin{figure}
\centering
\includegraphics[width=13.0cm]{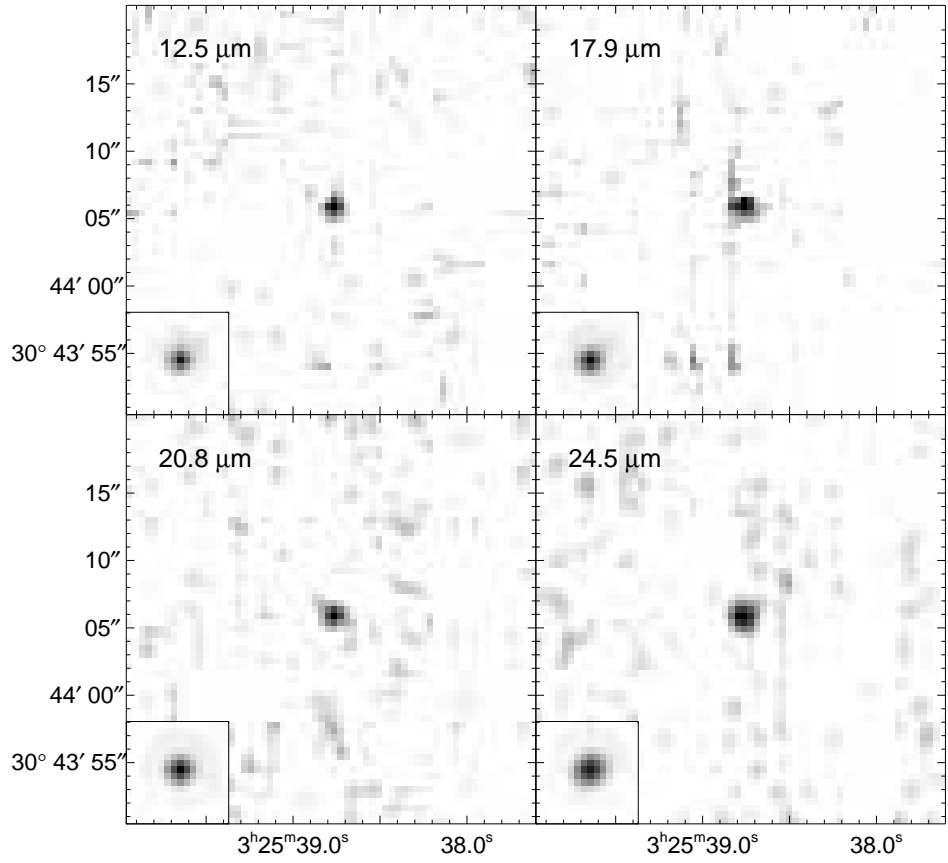}
\caption{{\bf MIRLIN images of L1448C:} 
Each panel is labeled with the corresponding
central wavelength of the filter used. The source is clearly detected
and indistinguishable from a point source at all wavelengths.  The 
image field-of-view is $30\farcs8 \times 30\farcs8$.\label{fig:imc}}
\end{figure}

\begin{figure}
\centering
\includegraphics[width=13.0cm]{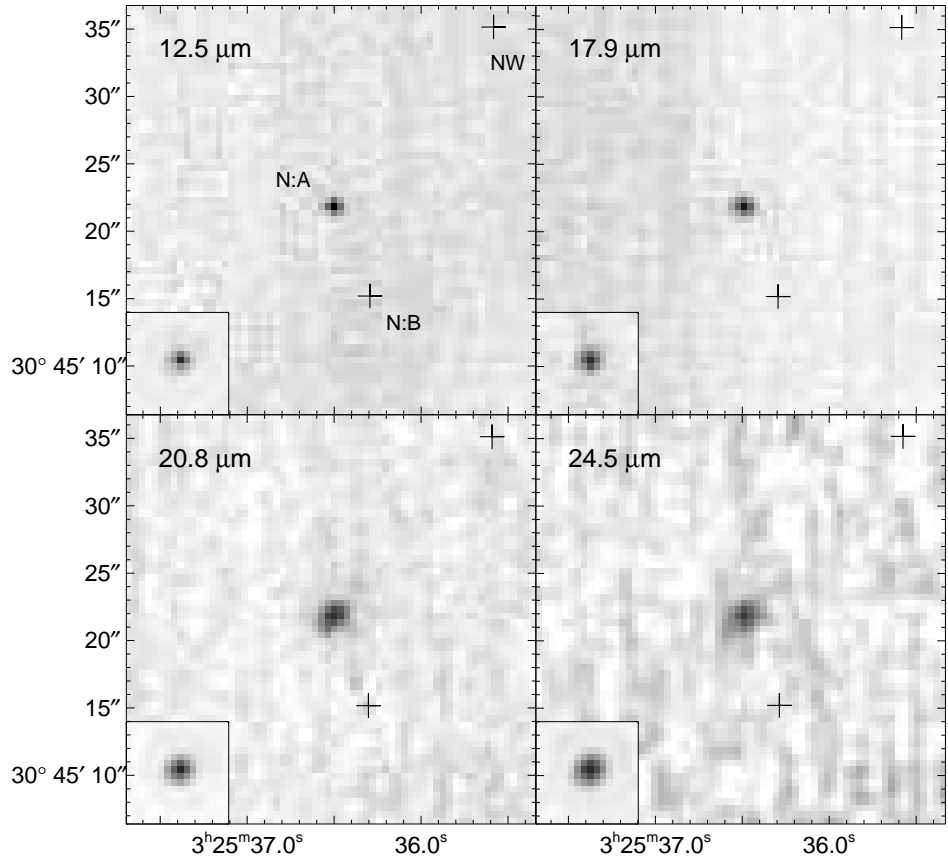}
\caption{{\bf MIRLIN images of L1448N:} 
Each panel is labeled with the corresponding
central wavelength of the filter used. The detected source is L1448N:A;
crosses indicate the positions of other YSOs (L1448N:B, L1448NW) in the field 
that were not detected. Faint elongations seen in these images are almost certainly
artifacts of the chop pattern. The image field-of-view is 
$30\farcs8 \times 30\farcs8$.\label{fig:imn}}
\end{figure}

Our observations of L1448 IRS~3 are presented in Figures \ref{fig:imc}
and \ref{fig:imn}; each of the
four panels is labeled with the central wavelength of the filter used, 
and has an image of the PSF at the
appropriate wavelength shown in the lower left corner.
Figure \ref{fig:imc} shows our observations of L1448C which,
at these resolutions ($0\farcs90$, $1\farcs29$, $1\farcs50$, and $1\farcs77$, 
at 12.5, 17.9, 20.8, and 24.5 \micron, respectively), 
is indistinguishable from an isolated point source.
Figure \ref{fig:imn} shows our images of the rest of the sources
in IRS~3, all of which lie within a single MIRLIN field of view:
the 7\arcsec\ binary, L1448N:A and B,
and a third source $\sim$20\arcsec\
away, L1448NW \citep{ter97,bar98,wo00a,ei00,lo00}.
Although there appears to be a hint of extended emission around L1448N:A in 
the 20.8 and 24.5 \micron\ images, the extension is north-west/south-east,
in good alignment with the direction of the chop, and therefore is most likely 
an artifact of the chop pattern.

In February 2000 we detected a single source in the 
L1448N/NW field, at the location of L1448N:A; this was initially somewhat
surprising, since N:B is actually quite 
a bit brighter than N:A at millimeter wavelengths.
We therefore made additional observations of this field in
October 2000.  Using the nearby bright star BS 999 as a position reference,
we find that the single source is consistent only with L1448N:A;
for it to be N:B or NW would require telescope positioning errors of
$7\farcs3$ and $17\farcs5$ (15 and 37 pixels), respectively.
In Figure \ref{fig:imn}, crosses show the positions at which N:B and
NW should appear.  These results agree with those presented by \citet{cia03} for
the L1448 N:A,B system.

\begin{figure}
\centering
\includegraphics[width=13.0cm]{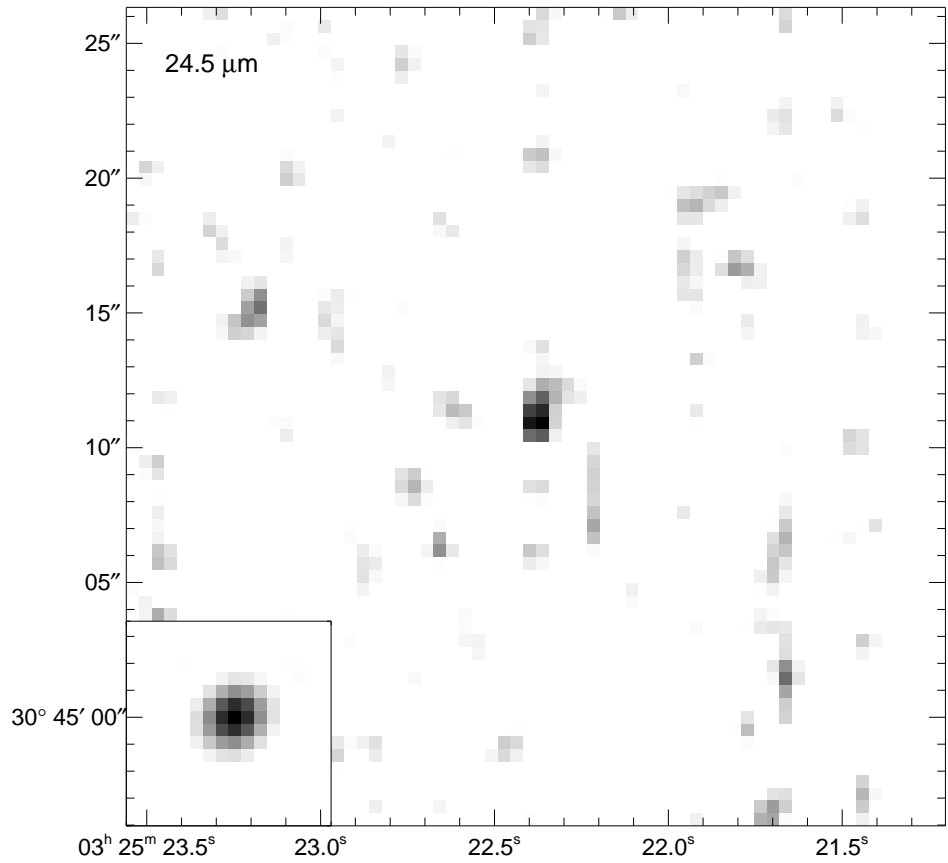}
\caption{{\bf MIRLIN image of L1448 IRS~2 at 24.5~\microns:} 
The source was marginally
detected at this wavelength, and undetected at 20.8~\microns.  We did
not attempt to observe it in any other filters.  The image field-of-view is
$30\farcs8 \times 30\farcs8$.\label{fig:irs2}}
\end{figure}

Figure \ref{fig:irs2} shows the MIRLIN image of L1448 IRS~2 at 24.5~\microns, 
along with the PSF at that wavelength.  IRS~2 was observed at 20.8 and 24.5~\microns, 
but detected (at 4$\sigma$) only at the longer wavelength.

Table 1 provides the coordinates of each source and
presents the MIRLIN photometry and upper limits. Coordinates for the
Class~0 sources L1448N:A, N:B and NW were obtained from \citet{lo00}; 
the positions of L1448C and L1448 IRS~2 were taken from \citet{an00}.
Errors in the photometry are dominated by the calibration uncertainties; 
even the most marginal detection, IRS 2, was 4$\sigma$.  For L1448N:A and 
L1448C, signal-to-noise ratios varied between 6 and 23.

\begin{figure}
\centering
\includegraphics[width=15.0cm]{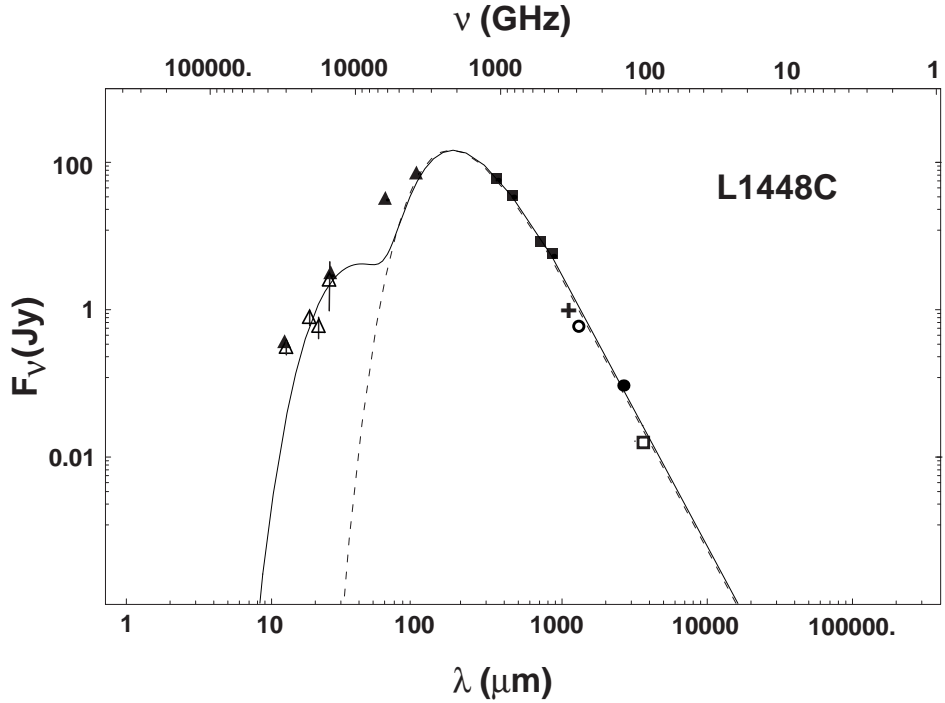}
\caption{{\bf Spectral energy distribution of L1448C:} 
The flux data for L1448C are presented, along with a
single-temperature greybody fit (dashed line), and a two-greybody
fit (solid line).  Some of the parameters of the dual-greybody fit
are presented in Table 2. The filled triangles represent 
{\em IRAS} photometry from 
HIRES point-source modeling \citep{bar98}; the open triangles with error bars
are from MIRLIN observations (this work). 
Filled squares present JCMT data from \citet{ch00}, the plus is an IRAM flux from
\citet{bar98}, the open circle is from \citet{bac91}.  The filled circle
indicates flux data from \citet{bac95}, while the open square is from \citet{gui92}.
\label{fig:sedc}}
\end{figure}

\begin{figure}
\centering
\includegraphics[width=15.0cm]{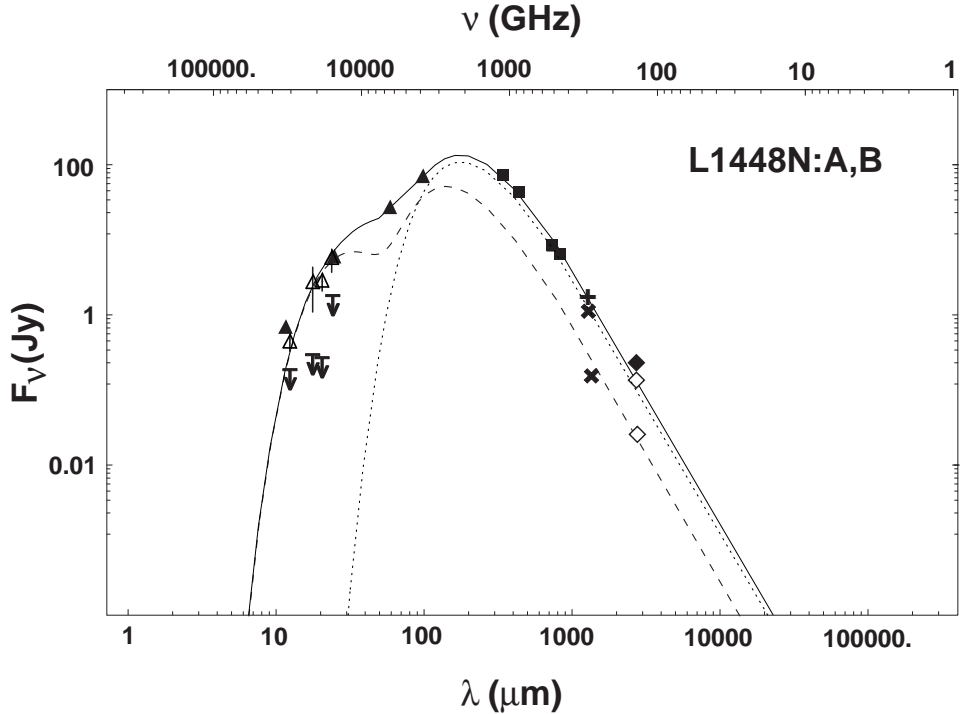}
\caption{{\bf Spectral energy distributions of the L1448N protobinary components:} 
The individual flux data for the two protobinary components, L1448N:A
and N:B are presented in conjunction with the literature flux values 
for the total emission from the L1448N binary at various wavelengths. The 
fits to the individual fluxes shown for each source 
(dashed and dotted lines correspond to fits for N:A and N:B, respectively) 
were generated assuming that each component emits approximately the
same flux at 100~\microns---(\emph{i.e.}, case 3: see \S\ref{sec:nab} for a discussion).
Some of the parameters of the single- and dual-greybody fits are presented in Table 2. 
The solid line represents the sum of the fits to the individual flux data
points (dashed + dotted).  The filled triangles represent {\em IRAS} photometry from 
HIRES point-source modeling \citep{bar98}; the open triangles with error bars and
the upper limits are from MIRLIN observations (this work). 
Filled squares indicate JCMT data from \citet{ch00}, the plus represents IRAM data from
\citet{bar98}.  The ``x''s indicate fluxes from \citet{ter97}.  The 
open diamonds are from \citet{lo00}, while the filled diamond is the sum 
of those fluxes.
\label{fig:sednab}}
\end{figure}

\begin{figure}
\centering
\includegraphics[width=15.0cm]{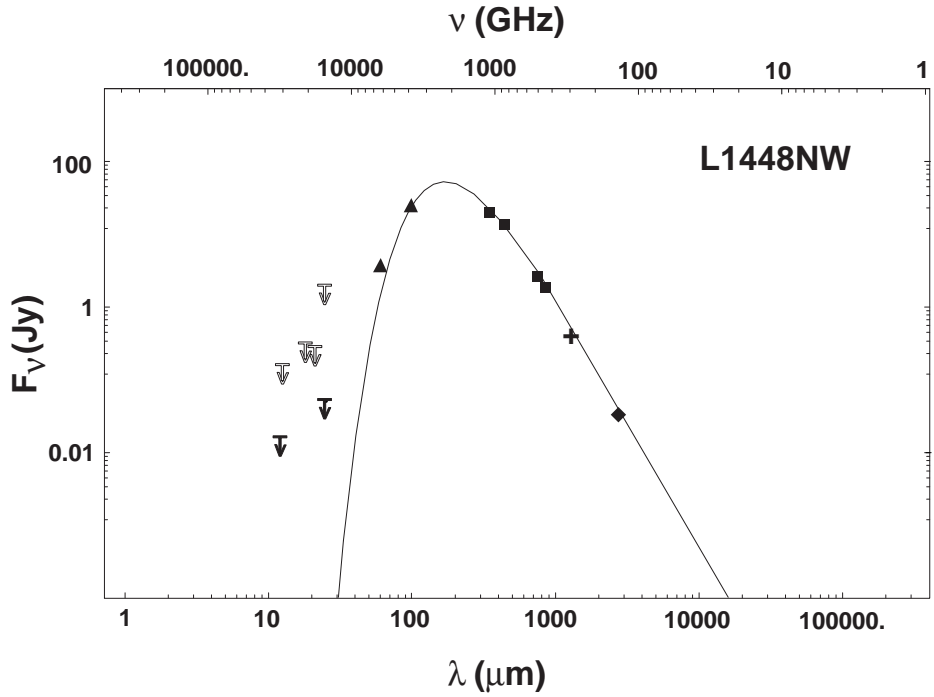}
\caption{{\bf Spectral energy distribution of L1448NW:} 
The flux data for L1448NW are presented, along with a single-temperature greybody fit
(solid line).  Some of the parameters of the single-greybody fit
are presented in Table 2.  The filled upper limits 
and the filled triangles represent {\em IRAS} photometry from HIRES point-source modeling 
\citep{bar98}; the open upper limits are from MIRLIN observations (this work). 
Filled squares indicate JCMT data from \citet{ch00}, the plus represents IRAM data from
\citet{bar98}, and the filled diamond is from \citet{lo00}.
\label{fig:sednw}}
\end{figure}

\begin{figure}
\centering
\includegraphics[width=15.0cm]{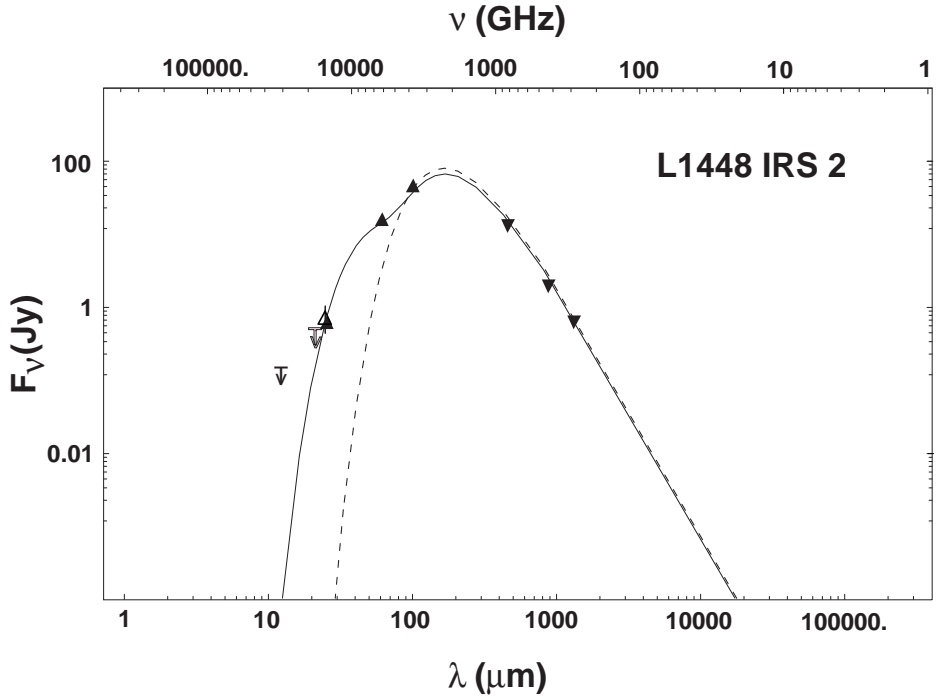}
\caption{{\bf Spectral energy distribution of L1448 IRS 2:} The
flux data for IRS 2 are presented, along with a
two-greybody fit (solid line) and a single-temperature greybody fit 
(dashed line). Some of the parameters of the dual-greybody fit
are presented in Table 2.  The filled upper limit and the filled triangles 
represent {\em IRAS} photometry from HIRES point-source modeling \citep{oli99};
the open triangle with the error bar and the open upper limit are MIRLIN data (this work). 
The filled inverted triangles indicate photometry obtained with SCUBA at the JCMT and 1.3mm 
continuum flux data from the 12-meter millimeter-wave dish at Kitt Peak \citep{oli99}.
\label{fig:sedi2}}
\end{figure}

Figures \ref{fig:sedc} -- \ref{fig:sedi2} present SED plots
for the three detected objects (L1448C, L1448N:A, and L1448 IRS~2), as well
as for the undetected sources, L1448N:B and L1448NW, constrained in the mid-infrared 
by the MIRLIN upper limits (indicated by arrows).  L1448NW, which is shown in
Figure \ref{fig:sednw} for completeness, was not detected and we
were unable to improve on previous upper limits in the mid-infrared that
were derived from HIRES point-source modeling \citep{bar98}.  Previously reported 
data from the literature are plotted in combination with our MIRLIN fluxes.
The SEDs are shown with either single or dual-greybody fits to the data.
Those SED figures with dual-greybody fits (solid lines) also 
display single-temperature greybodies as dashed lines, each greybody being a modified 
blackbody of the form:

\begin{equation}
S_{\nu}\ =\ B_{\nu}(T_d)(1\ -\ e^{-\tau_{\nu}})\,d\Omega,
\end{equation}

\noindent where $B_{\nu}(T_d)$ is the Planck function at frequency $\nu$
and dust temperature T$_d$, $\tau_{\nu}$ is the dust optical depth, and
$d\Omega$ is the solid angle subtended by the source.  The sole exception to
this convention is Figure \ref{fig:sednab}, which presents fits for the data from 
each component of the L1448N protobinary using a dashed line for N:A and a dotted 
line for N:B, while the solid line represents the sum of those two individual SED fits; note
that this particular pair of fits to the data is {\em not} the only possible solution, and this 
fact is discussed at length in \S\ref{sec:nab}.  

The various dual-greybody model fits to the data, shown by the solid lines in 
Figures \ref{fig:sedc} -- \ref{fig:sedi2}, were produced by simply summing together 
two greybody curves, each with a different effective dust temperature. 
It is clearly not possible to fit all of the data points for any of the sources 
which were detected by MIRLIN in the mid-infrared with a single-temperature greybody. 
The dual-greybody curves yield more accurate calculations of the bolometric luminosities. 
The model fits to the flux data for each source and any necessary assumptions are discussed 
in detail in \S\ref{sec:disc}.

Table 2 lists each source, the number of greybodies
(N$_{gb}$) used to fit the flux data, the associated dust
temperatures (T$_{d}$) for each greybody, and
two physical parameters derived for these sources
from the SED fitting:
the bolometric luminosity, L$_{bol}$, and
the infall envelope mass, M$_{env}$, estimated from \citep{hi83}:

\begin{equation}
M_{env}\ =\ {S_{\lambda}d^2 \over \kappa_{\lambda}B_{\lambda}(T_d)}
\end{equation}
evaluated at $\lambda = 1.3$ mm.

Recall that dust emission is generally optically thin at millimeter wavelengths, 
and is therefore a direct indicator of the \emph{total} mass 
(gas $+$ dust) contained in molecular cloud cores, given appropriate assumptions 
about the dust--to--gas ratio.  The appropriate value 
of $\kappa_{1.3\;\mbox{mm}}$ to use depends on the status or ``age'' of the 
YSOs in the core in question, due to the evolution of dust properties \citep{an00}.
We used $\kappa_{1.3\;\mbox{mm}} = 0.01$ based on literature values for 
similar regions \citep{an93,an94}. 

In addition, the ratio of the submillimeter luminosity to the bolometric luminosity,
L$_{submm}$/L$_{bol}$, where L$_{submm}$ is calculated using wavelengths
longer than 350~\microns, was determined for each source and included in 
Table 2. This ratio can be used as a diagnostic tool to determine 
the evolutionary status of embedded YSOs, because it is linked with the ratio
of the circumstellar envelope mass to the mass of the central accreting
core, and thus decreases with time. A value of L$_{submm}$/L$_{bol}$ $>$
0.5\% indicates M$_{env}$/M$_*$ $>$ 1, {\em i.e.} an envelope mass greater than
the protostellar core mass, a distinguishing characteristic of Class~0
sources (see \S\ref{sec:intro}).

\section{Discussion}\label{sec:disc}

\subsection{Spectral Energy Distributions - Individual Sources}

\subsubsection{L1448C}

This Class~0 source, which drives a powerful, highly-collimated outflow,
is a favorite target of many observers \citep[see][and references
therein]{bar98}. When first discovered, it was believed to be one of the
youngest Class~0 sources found, with a kinematic age (based on the outflow
parameters) of $\sim$3500 years \citep{bac90}. This figure was revised
upwards to $\sim$32,000 years by \citet{bar98}, who found that the outflow
was far more extensive than previously thought \citep[see also][]{wo00a}.

The SED for L1448C is presented in Figure \ref{fig:sedc}, with a
two-greybody fit to the data shown by the solid line. The dashed line
indicates a single-temperature greybody fit to the data.  

\subsubsection{L1448N:A,B}\label{sec:nab}

In order to analyze the SEDs for L1448N:A and N:B, and use them to derive
the parameters listed in Table 2, it is necessary to make
certain assumptions regarding the individual fluxes of these two objects at
far-infrared wavelengths. There are no high-angular-resolution data that
can separate the 7\arcsec\ protobinary components at or near the peak of
the dust emission ($\sim$100~\microns), and this causes problems when
attempting to fit modified blackbodies to the spectral energy distributions
of these objects.

Due to the excellent correspondence between the 12 and 25~\micron\ {\it
IRAS} HIRES point-source model fluxes from \citet{bar98} for L1448N (which represent 
the sums of the fluxes for both components of the protobinary) and our MIRLIN
fluxes (see Figure \ref{fig:sednab}), it seems clear that the mid-infrared
fluxes are due chiefly to L1448N:A, with very little, if any, contribution
from N:B or NW.  Again, these results agree with those presented in \citet{cia03}. 
In the millimeter, the situation is quite different: N:B
produces most of the flux seen at 2.7 mm \citep{ter97,lo00}. Using HIRES
modeling, \citet{bar98} showed it is possible to separate the {\it
IRAS} fluxes produced by L1448NW from the total flux emitted by the L1448N
triple system. The L1448N system produces a total of 89~Jy at 100~\microns,
while the HIRES-modeled 100~\micron\ flux from NW is $\sim$23~Jy. Therefore the {\it
maximum} possible flux from the protobinary at 100~\microns\ must be on the
order of 66~Jy. Since some of the emission detected by \emph{IRAS} at that
wavelength is presumably due to shock-heated dust caused by the impact of the L1448C
outflow on the dense core containing the L1448N protobinary, the true
100~\micron\ flux from the binary is probably somewhat less than 66~Jy
\citep{bar98,cur99}.
 
The flux ratio is approximately 5:1 in favor of N:B in the millimeter regime
\citep{lo00}, therefore it seems reasonable to explore a range of
possibilities for the 100~\micron\ fluxes.  We have done so, in the following fashion: 
(1) Assume that the fluxes at 100~\microns\ have the same ratio as the 2.7 mm fluxes, so that 
N:A produces 11~Jy and N:B emits the remaining 55~Jy; (2) assume the inverse scenario 
(\emph{i.e.}, N:B has a flux of 11~Jy, N:A produces the other 55~Jy).  These
two possibilities define the endpoints of the range that we consider in this paper.
A third instructive case to consider is the intermediate scenario which involves dividing 
the 66~Jy into equal portions; this assumes fluxes of $\sim$33~Jy at 100~\micron\ from
each of the protobinary components.

The assumptions inherent in this range of possibilities were used to generate 
a series of different SEDs for N:A and N:B; each of the SEDs for N:A were then fit 
with two-greybody models, while those for N:B were done with single-temperature
greybodies. 
Sanity checks were performed for each case by using the various fits to calculate
ranges of L$_{bol}$ for the two sources, and thus to ascertain
whether or not the sum of their luminosities approximates the total luminosity 
expected for the protobinary. The luminosity of the protobinary was derived by 
taking the total fluxes of the L1448N system reported in the literature and
subtracting the flux data for L1448NW at all wavelengths, then plotting the
resultant SED and fitting with the sum of two modified blackbodies; this
yields a maximum expected L$_{bol}$ of $\sim$14.1~L$_{\sun}$ for the
protobinary system. The derivation of the L1448NW luminosity, using a
single greybody fit to the data, is discussed in \S\ref{sec:nw}.

From the analysis described above, we find the plausible
range of L$_{bol}$ for L1448N:A is 3.1--6.6~L$_{\sun}$, and for N:B is
2.8--7.4~L$_{\sun}$.  Since N:B dominates in the millimeter, and N:A in the 
mid-infrared, and the two objects have bolometric luminosities of
the same order of magnitude, their SEDs must cross each other somewhere
in the far-infrared, probably in the general vicinity of 100~\microns.
The scenario of approximately equal 100 \micron\ fluxes thus seems the most 
illustrative; it yields bolometric luminosities of $\sim$4.8~L$_{\sun}$ for N:A, 
and 5.5~L$_{\sun}$ for N:B.  The sum is indeed less than the maximum expected 
luminosity.

Figure \ref{fig:sednab} is a plot of the literature and MIRLIN flux data for each
protobinary component along with a two-greybody fit for L1448N:A (dashed line) and
a single-temperature greybody fit for N:B (dotted line) using the ``equal 100 \micron\ fluxes'' 
scenario. The previously published data points and derived fluxes for the entire
L1448N protobinary system are plotted, and the sum of the fits to the individual 
SEDs (i.e., dashed + dotted) is shown as a solid line. The values of the physical 
parameters derived from the assumption of approximately equal 100 \micron\ fluxes 
(case 3) are reported in Table 2.

\subsubsection{L1448NW}\label{sec:nw}

L1448NW was not detected by MIRLIN at any wavelength, as may be seen in
Figure \ref{fig:imn}. This was the expected result, due to the extensive
HIRES point-source modeling work reported in \citet{bar98}, which
established \emph{IRAS} upper limits for this low-luminosity source of 0.015
and 0.05~Jy at 12 and 25~\microns, respectively. Our MIRLIN upper limits
do not improve on these results (see Table 1).  

We have incorporated the submillimeter and millimeter-wave flux data 
from \citet{ch00} and \citet{lo00} in the plot shown in
Figure \ref{fig:sednw}; these data are entirely consistent with the
FIR \emph{IRAS} fluxes derived from HIRES modeling \citep{bar98}. The
results from a single-temperature greybody fit to the data
at wavelengths longer than 60~\microns\ are used to calculate the parameters 
reported in Table 2, which allow us to place L1448NW on the
evolutionary diagram in Figure \ref{fig:evo}.

\subsubsection{L1448 IRS~2}\label{sec:irs2}

L1448 IRS~2 was identified as a Class~0 source by \citet{oli99}, and first
proposed as a candidate protobinary in \citet{wo00a}, based on CO
observations which mapped two distinct molecular outflows driven by IRS~2.
\citet{vo02} claimed detection of both binary components using BIMA,
although the spatial separation was not reported.
With the high spatial resolution of MIRLIN at the IRTF ($<$2\arcsec\ at
all wavelengths), we had hoped to be able to detect and resolve both
components of the system, but only a single source was found,
at the 4$\sigma$ level in the 24.5~\micron\ filter (see Figure \ref{fig:irs2}).
This suggests that either the binary components are too close together for MIRLIN to 
resolve (see the first paragraph of \S\ref{sec:res} for MIRLIN resolutions), 
or that the fluxes from one of the components are below the detection thresholds at 
all MIRLIN wavebands.

Figure \ref{fig:sedi2} presents the SED for IRS~2, with a two-greybody fit
to the data shown by the solid line. The dashed line
indicates a single-temperature greybody fit to the data.

\subsection{Evolutionary Status of L1448 Sources}

\subsubsection{M$_{env}$ vs. L$_{bol}$}

\begin{figure}
\centering
\includegraphics[width=15.0cm]{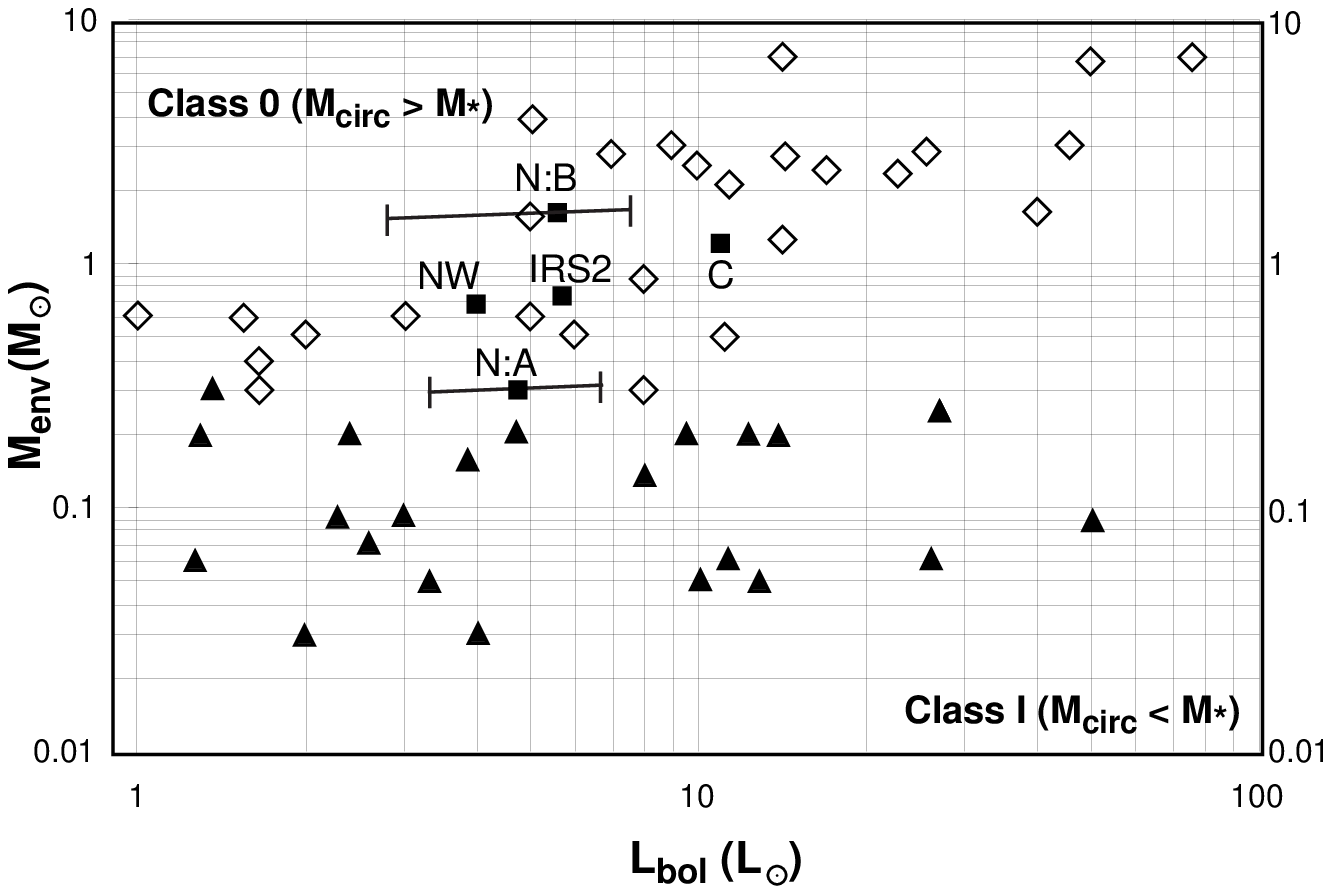}
\caption{{\bf Evolutionary status diagram of the L1448 objects--
M$_{env}$ vs. L$_{bol}$:}  We plot the positions of the L1448 protostars (filled squares)
along with a number of other well-known Class 0 (open diamonds) and Class I sources (filled
triangles) from \citet{an00} and \citet{bon96} in the circumstellar mass vs. bolometric 
luminosity diagram. Since bolometric luminosity
is an indirect measure of central core mass, this is essentially a plot
of envelope mass vs. central core mass.  The ``error bars''
on L1448N:A and N:B represent the full range of likely values for
each object (all three cases, with the filled square being case 3,
see \S\ref{sec:nab}).  Generally, sources to the upper left of the 
diagram are in the Class 0 region, while sources to the lower right 
are Class I objects.\label{fig:evo}}
\end{figure}

Figure \ref{fig:evo} is a plot of M$_{env}$ vs. L$_{bol}$ for all of the
sources in this paper (filled squares), along with all other Class 0 (open diamonds) 
and Class I YSOs (filled triangles) with L$_{bol} \> 1.0$ from Table 1 in \citet{bon96} 
plotted for comparison.  This plot may be used as a diagnostic tool for determining the
relative ages of embedded sources \citep{an94,sar96}. The diagram shows
infall envelope mass (along the vertical axis) vs. bolometric luminosity
(horizontal axis). L$_{bol}$ is directly related to the mass of the central
protostellar core at very early stages of evolution, assuming all
luminosity is generated by gravitational infall, according to:

\begin{equation}
L_{bol}\ =\ {G{\dot M} M_* \over R_*},
\end{equation} 

\noindent where a protostellar radius of $\sim$3~R$_{\sun}$ is assumed \citep{sta80b}.

Our results depicted in Figure \ref{fig:evo} for L1448C and L1448NW are
consistent (within errors) with those presented in \citet{bar98} for those two 
sources. The new results in the diagram presented here are (1) L1448 IRS~2 is
plotted on the diagram to allow comparison with the other sources from this
cloud, and (2) we have plotted the L1448N protobinary components {\it
separately} on this graph, using the assumptions discussed in \S\ref{sec:nab}.
 
All of the target objects discussed in this paper lie within
the Class~0 region of the evolutionary diagram in Figure \ref{fig:evo}, although 
L1448N:A appears to be close to the transition zone between 
Class~0 and Class I, indicating N:A may be the most evolved source in this cloud 
after the Class I source L1448 IRS 1.  By the same criteria, L1448N:B is very
likely the youngest source that we attempted to image in this cloud.  These 
classifications are reflected in the values of the parameter L$_{submm}$/L$_{bol}$, 
listed in Table 2, which range from 1.7--7.3\%: the value of 1.7\%
corresponds to L1448N:A, and 7.3\% to N:B (for the case 3 scenario of equal 
100 \micron\ fluxes).

We note that recently \citet{cia03} have argued that L1448N:A is probably
a Class I object. However, our smallest value of L$_{submm}$/L$_{bol}$ for N:A,
derived for case 2 (assuming 100 \micron\ flux for N:A $\sim 55$ Jy, for N:B $\sim 11$ Jy),
was 1.5\%, which lies well above the lower-limit value of 0.5\% established for 
Class 0 objects by \citet{an00}.  We point out two additional factors which bolster 
the argument for the Class 0 status of N:A:
(1) Its molecular outflow is extremely well-collimated, similar to outflows
associated with other Class 0 sources \citep{wo00a}.
(2) We find that it is not detected in the near-infrared by 2MASS, although the 
Class I source L1448 IRS 1 is strongly detected in
all 2MASS bands (JHK$_s$). Although fluxes are presented in all bands
for a ``source'' detected by 2MASS within a few arcseconds of L1448N:A,
the J-band flux is presented as an upper limit, and the H-band and K-band
fluxes suffer photometric confusion. The 2MASS Quicklook images
reveal a bow-shaped knot north of L1448N:A in all three bands, and
an additional jet-like feature south of L1448N:A in the K-band image. 
Inspection of these features reveals that they correspond to 
H$_2$ outflow features imaged by \citet{ei00}. The jet-like feature
in the K-band image of L1448N:A is part of the blueshifted outflow lobe
associated with L1448C, which is located $\sim 1.5^{\prime}$ to the
southeast of the L1448N sources. This outflow was also identified in
HIRES-processed {\it IRAS} images and high-velocity CO emission \citep{bar98,wo00a} 
The bow-shaped knot that appears just north
of L1448N:A in the J,H, \& K$_s$ 2MASS images is very prominent in the H$_2$
image. There is no trace of a near-infrared point source counterpart to L1448N:A in
the 2MASS data.

Additional near-infrared data for this source became available during a late revision of this paper: 
a FLAMINGOS survey of L1448 down to $m_{K_s} \sim$ 17 mag \citep{tsu05}.  Again, no 
near-IR counterpart to L1448N:A was found during this survey.  However, these same investigators
also used {\em Chandra} to observe L1448; they detected a weak X-ray source in the 
general vicinity of N:A.  No X-ray emission was detected from any of the other
Class 0 sources under discussion in this paper.  \citet{tsu05} discuss the fact that, given the non-detection
in the near-infrared bands, L1448N:A must have an extremely steep near-to-mid-infrared spectral index
($\alpha>3.2$), and that nearly all known Class I protostars have $\alpha \leq 2$ (the single
exception being the peculiar young stellar object, WL 22, with $\alpha = 3$).

We therefore contend that all currently available data for L1448N:A indicate that it is a bonafide
(albeit borderline) Class 0 source; the weak X-ray detection does not rule this out, and
all other data support this conclusion.

\subsubsection{Effects of the L1448C Outflow on the L1448N/NW Core}\label{sec:out}

Using ammonia data, \citet{cur99} report that the high-velocity L1448C outflow has 
had a significant affect on the northern core, creating a deep ``dent'' in its southern 
edge, directly south of the protobinary. \citet{bar98} suggest that the impact of the 
outflow from L1448C on the dense core fragmented it and actually induced the formation 
of the L1448N/NW system.

At first glance, this scenario of outflow-induced fragmentation appears to be at odds with 
our findings above that N:A is the most evolved Class 0 source in this cloud, while N:B is 
the youngest and most deeply embedded.  Of course, if these two sources do not form a 
gravitationally-bound protobinary system, it is unnecessary for them to be coeval. 
But for the purposes of this discussion, we will assume with \citet{ter97}, \citet{bar98},
\citet{ei00}, and 
\citet{wo00a}, that N:A and N:B are the bound components of a protobinary. Is it possible 
for presumably coeval sources, which appear to be physically similar, to travel different 
evolutionary paths?

One recently-proposed explanation for such a dichotemy in evolutionary status of protobinary 
components is found in the disintegration of unstable triple or 
higher-order multiple systems; such events may cause disk truncation for some of the 
components, subsequent episodic outflow activity linked to giant Herbig-Haro flows, and accelerated 
evolution of one or more components due to the rapid dissipation of circumstellar envelopes 
\citep{rei00, rei01}. 
It is true that most of the young sources in L1448 have been identified as driving sources of 
parsec-scale Herbig-Haro flows \citep{bal97,bar98,oli99,ei00,wo00a}. However, at 
$\sim$2000~AU separation, with a 
rotational period on the order of 60,000 years \citep{ter97}, the L1448N:A,B system does 
not seem to fit into this particular theoretical paradigm, since a remnant binary from
such a former higher-multiplicity system is predicted to have a much smaller separation. The 
L1448N:A,B components appear to be too young to have had time to interact with 
each other or with L1448NW in any significant way.  Furthermore, according to the \citet{rei00} 
theory, binary components that are remnants of unstable multiples having undergone ejection 
events might be expected to have more pronounced differences between their apparent evolutionary 
stages; \emph{e.g.}, one component may transition abruptly to Class II status, while the other 
remains in Class 0 or Class I.  All of the L1448 sources under discussion in this paper still fall 
within the same evolutionary category, although N:A appears to be close to transitioning to Class I.

We propose an alternative explanation for the (admittedly minor) difference between the protobinary
components: this could be nothing more than a chance side effect of the powerful L1448C outflow.  
Lacking a good understanding of the three-dimensional relative geometry of the sources in this 
cloud, we are unable to verify our hypothesis, but one possible scenario is that the outflow 
from L1448C has partially stripped the envelope surrounding N:A, while leaving N:B deeply embedded.  
This would allow N:A and N:B to be coeval sources that formed due to the impact of the L1448C 
outflow, while explaining the different \emph{apparent} ages of these objects.  

\section{Summary}

We have conducted a mid-infrared imaging survey of known Class~0 protostars
in L1448 with MIRLIN at the IRTF, using the 12.5, 17.9, 20.8, and
24.5~\micron\ filters. Of the five confirmed Class~0 sources in this cluster,
we imaged and obtained photometry for three, including a single source that 
was marginally detected at 24.5~\micron\ at the location of the
protobinary in L1448 IRS 2. We report upper limits for two undetected objects.

SEDs are presented for all five sources, and are used to derive various physical properties
of these sources, allowing us to place them on an evolutionary status diagram.  The components 
of the protobinary system L1448N are plotted separately; 
we find that the source known as L1448N:A lies significantly closer to the Class I portion of
the diagram than does its more deeply embedded companion, N:B.  One possible
explanation for this difference in apparent ``age'' may be found in the ``stripping'' of the
envelope surrounding N:A by the L1448C outflow, which is known to be colliding with the
northern core.

\acknowledgements 

We thank the IRTF support staff for assistance with the data acquisition.
JO acknowledges financial support by NASA Grants to the Wide-Field Infrared
Explorer Project and the Space Infrared Telescope Facility Project at the
Jet Propulsion Laboratory, California Institute of Technology. 

Many thanks to Michael Werner, Vincent Mannings, and our anonymous referee for 
helpful suggestions which improved this paper immensely.

This research has made extensive use of the Abstract Service maintained by the
NASA Astrophysics Data System, as well as the Simbad and VizieR databases archived
at CDS, Strasbourg, France.

This publication makes use of data products from the Two Micron All Sky Survey, 
which is a joint project of the University of Massachusetts and the Infrared Processing 
and Analysis Center/California Institute of Technology, funded by the National Aeronautics 
and Space Administration and the National Science Foundation.

This research has made use of the NASA/ IPAC Infrared Science Archive, which is operated 
by the Jet Propulsion Laboratory, California Institute of Technology, under contract with 
the National Aeronautics and Space Administration.

Portions of this work were carried out at the Jet Propulsion Laboratory,
California Institute of Technology, under contract with the National
Aeronautics and Space Administration. Development of MIRLIN was supported
by the JPL Director's Discretionary Fund and its continued operation is
funded by an SR+T award from NASA's Office of Space Science.

\clearpage
%
% 1st figcaption is location figure

\clearpage

% 2nd figcaption is for the L1448C images at N5, Qs, Q3, Q5

\clearpage

% 3rd figcaption is for the L1448N images at N5, Qs, Q3, Q5

\clearpage

% 4th figcaption is for the L1448 IRS~2 image at Q5

\clearpage

% 5th figcaption is SED figure for L1448C

\clearpage

% 6th figcaption is SED figure for L1448N:A,B

\clearpage

% 7th figcaption is SED figure for L1448NW

\clearpage

% 8th figcaption is SED figure for L1448 IRS 2

\clearpage

% 9th figcaption is for the Evolutionary Diagram

\clearpage

\begin{deluxetable}{lrclrclccccc}
\tabletypesize{\scriptsize}
%
%Table 1 should have 11 columns
%
\tablenum{1}
\tablewidth{0pc} 
\tablecaption{Source List \& MIRLIN Photometry$^{\dag}$}
\tablehead{
\colhead{Source} &\multicolumn{5}{c}{Coordinates (J2000.0)}& \colhead{} 
& \colhead{F$_{12.5}$} & \colhead{F$_{17.9}$}  & \colhead{F$_{20.8}$} 
& \colhead{F$_{24.5}$}  \\
\colhead{} & \multicolumn{3}{c}{Right Ascension}&\multicolumn{3}{c}{Declination}&
\colhead{} & \colhead{} &\colhead{} & \colhead{}  \\
\colhead{} & \colhead{h} & \colhead{m} & \colhead{s} &
\colhead{\arcdeg} & \colhead{\arcmin} & \colhead{\arcsec} &
\colhead{(Jy)} &\colhead{(Jy)} &\colhead{(Jy)} &\colhead{(Jy)}  \\
}
\startdata
L1448 IRS~2 & 03 & 25 &  22.4 & $+$30 &   45 & 12 & \nodata & \nodata & $<$0.46 & 0.64$\pm$0.23 \\
L1448NW & 03 & 25 &  35.653 & $+$30&   45& 34.20 & $<$0.15 & $<$0.30 & $<$0.27 & $<$1.85 \\
L1448N:B  & 03 & 25 &  36.339 &  $+$30 &  45 & 14.94 &  $<$0.15 &  $<$0.30 & $<$0.27 & $<$1.85 \\
L1448N:A  & 03 & 25 &  36.532 &  $+$30 &  45 & 21.35 &  0.43$\pm$0.10 & 2.62$\pm$1.54 & 2.80$\pm$0.65 & 5.44$\pm$1.80 \\
L1448C  & 03&  25&  38.8& $+$30&   44&    05.0 &  0.29$\pm$0.04 &  0.73$\pm$0.14  & 0.55$\pm$0.17 & 2.37$\pm$1.52 \\
%
% Text for table footnotes follows the tabular data BEFORE THE \enddata
% statement (MB) and must be inside the
% deluxetable environment.  Note that it is OK to put \ref's in 
% \tablenotetext's.
% 
%\tablenotetext{a}{Sample footnote for table~\ref{tbl-1} that was generated
% with the deluxetable environment}
%\tablenotetext{b}{Another sample footnote for table~\ref{tbl-1}}
%\tablenotetext{c}{Footnote with no call out}
%\tablenotetext{d}{Another footnote with no call out}
%\tablenotetext{e}{A further additional footnote with no call out}
%
\tablenotetext{\dag} {All upper limits are 3$\sigma$.}
\enddata \label{tab:phot}
\end{deluxetable}

\begin{deluxetable}{lccccccc} 
\tabletypesize{\scriptsize}
\tablenum{2}
\tablecolumns{6} 
\tablewidth{0pc} 
\tablecaption{Derived Physical Parameters} 
\tablehead{ 
\colhead{Source}
& \colhead{N$_{gb}$}
& \colhead{T$_{d}$ (K)} 
& \colhead{L$_{bol}$ (L$_{\sun}$)}  
& \colhead{M$_{env}$ (M$_{\sun}$)}
& \colhead{L$_{submm}$/L$_{bol}$ (\%)}}
\startdata 
L1448 IRS~2  & 2 &    20,50     &  5.6 & 0.73 & 3.3  \\
L1448NW      & 1 &    20     &  3.5 & 0.68 & 5.0  \\
L1448N:B  \dag   & 1 &    17.5     &  5.5 & 1.75 & 7.3  \\
L1448N:A  \dag   & 2 &    23,90     &  4.8 & 0.29 & 1.7  \\
L1448C       & 2 &   19,85 & 10.0 & 1.10 & 3.7  \\
\tablenotetext{\dag} {Parameters for L1448N:A,B are derived by assuming approximately equal 100 \micron\ fluxes.}
\enddata \label{tab:phys}
\end{deluxetable} 


\newpage

\begin{thebibliography}{}

\bibitem[Andr\'{e}, Ward-Thompson, and Barsony(1993)]{an93} Andr\'{e}, Ph.,
Ward-Thompson, D., and Barsony, M.  1993, \apj, 406, 122

\bibitem[Andr\'{e} \& Montmerle(1994)]{an94} Andr\'{e}, Ph., and Montmerle, T.  1994, 
\apj, 420, 837

\bibitem[Andr\'{e}, Ward-Thompson, and Barsony(2000)]{an00} Andr\'{e}, Ph.,
Ward-Thompson, D., and Barsony, M. 2000, in Protostars and Planets IV, 
ed. V.~Mannings, A.~P.~Boss, and S.~S.~Russell 

\bibitem[Anglada et al.(1989)]{ang89} Anglada, G., Rodr\'{i}guez, L., Torrelles, J., Estalella, R.,
Ho, P., Cant\'{o}, J., L\'{o}pez, R., \& Verdes-Montenegro, L. 1989, \apj, 341, 208

\bibitem[Aumann, Fowler, and Melnyk(1990)]{aum90} Aumann, H.~H., Fowler, J.~W., 
and Melnyk, M. 1990, \aj, 99, 1674   
 
\bibitem[Bachiller \& Cernicharo(1986a)]{ba86a} Bachiller, R. and Cernicharo, J.
1986a, \aap, 166, 283 

\bibitem[Bachiller \& Cernicharo(1986b)]{ba86b} Bachiller, R. and Cernicharo, J.
1986b, \aap, 168, 262 

\bibitem[Bachiller\ et al.(1990)]{bac90} Bachiller, R., Cernicharo, J.,
Martin-Pintado, J., Tafalla, M., and Lazareff, B.  1990, \aap, 231, 174 

\bibitem[Bachiller, Andr\'{e}, \& Cabrit(1991)]{bac91} Bachiller, R., Andr\'{e}, P., \&
Cabrit, S. 1990, \aap, 241, L43 

\bibitem[Bachiller\ et al.(1995)]{bac95} Bachiller, R., Guilloteau, S., Dutrey, A.,
Planesas, P., \& Mart\'{i}n-Pintado, J.  1995, \aap, 299, 857

\bibitem[Bachiller\ et al.(2000)]{bac00} Bachiller, R., Gueth, F., Guilloteau, S., 
Tafalla, M., \& Dutrey, A., 2000, \aap, 362, L33

\bibitem[Bally\ et al.(1997)]{bal97} Bally, J., Devine, D., Alten, V., \& Sutherland, R. 1997,
\apj, 478, 603

\bibitem[Barsony(1994)]{bar94} Barsony, M.  1994, in Clouds, Cores and Low Mass
Stars, ed. D.~P.~Clemens \& R.~Barvainis, ASP Conference Series, 65, 197

\bibitem[Barsony et al.(1998)]{bar98} Barsony, M., Ward-Thompson, D., Andr\'e,
P., and O'Linger, J. 1998, \apj, 509, 733

\bibitem[Barsony(2000)]{bar00} Barsony, M. 2000, private communication

\bibitem[Bontemps et al.(1996)]{bon96} Bontemps, S., Andr\'e, P., Terebey, S., 
and Cabrit, S. 1996, \aap, 311, 858

\bibitem[Carkner(1998)]{car98} Carkner, L. 1998, Ph.D. Dissertation, Pennsylvania State University

\bibitem[Chandler \& Richer(2000)]{ch00} Chandler, C., \& Richer, J. 2000, \apj, 530, 851

\bibitem[Chen et al.(1995)]{che95} Chen, H., Myers, P., Ladd, E., and Wood, D. 1995,
\apj, 445, 377

\bibitem[Chen et al.(1997)]{che97} Chen, H., Grenfell, T., Myers, P., and Hughes, J. 1997,
\apj, 478, 295

\bibitem[Ciardi et al.(2003)]{cia03} Ciardi, D., Telesco, C., Williams, J., Fisher, R.,
Packham, C., Pi\~na, R., Radomski, J.  2003, \apj, 585, 392

\bibitem[Cohen \& Kuhi(1979)]{co79} Cohen, M., \& Kuhi, L. 1979, \apjs, 41, 743

\bibitem[Curiel et al.(1999)]{cur99} Curiel, S., Torrelles, J., Rodr\'{i}guez,
L.~F., G\'{o}mez, J., \& Anglada, G. 1999, \apj, 527, 310

\bibitem[Eisl\"{o}ffel(2000)]{ei00} Eisl\"{o}ffel, J. 2000, \aap, 354, 236

\bibitem[Froebrich(2005)]{fro05} Froebrich, D. 2005, \apjs, 156, 169

\bibitem[Greene et al.(1994)]{gre94} Greene, T., Wilking, B., Andr\'e, P., and Young, E.
1994, \apj, 434, 614

%\bibitem[Gregersen et al.(1997)]{gre97} Gregersen, E., Evans, N., Zhou, S., and Choi, M. 
%1997, \apj, 484, 256

\bibitem[Guilloteau\ et al.(1992)]{gui92} Guilloteau, S., Bachiller, R., Fuente, A., 
Lucas, R.  1992, \aap, 265, L49

\bibitem[Hildebrand(1983)]{hi83} Hildebrand, R. 1983, QJRAS, 24, 267

\bibitem[Hurt \& Barsony(1996)]{hur96} Hurt, R., and Barsony, M. 1996, \apj, 460, L45 

\bibitem[Lada \& Wilking(1984)]{lad84} Lada, C.~J. \& Wilking, B.~A.  1984, \apj, 287, 610 

\bibitem[Lada(1987)]{lad87} Lada, C.~J. 1987, in {\em I.~A.~U.~Symposium No.~115: 
Star Forming Regions}, eds.  M.~Peimbert \&  J.~Jugaku, (Dordrecht: Reidel), 1

\bibitem[Lada(1991)]{lad91} Lada, C. 1991, in {\em The Physics of Star Formation and 
Early Stellar Evolution}, ed. C.~J.~Lada \& N.~D.~Kylafis, NATO ASI Series C, 342, 329

\bibitem[Ladd et al.(1991)]{la91} Ladd, E., Adams, F., Casey, S., Davidson, J., Fuller, G.,
Harper, D., Myers, P., and Padman, R. 1991, \apj, 366, 203 

%\bibitem[Larionov et al.(1999)]{lar99} Larionov, G., Val'tts, I., Winnberg, A., Johansson, L.,
%Booth, R., and Golubev, V. 1999, \aaps, 139, 257

%\bibitem[Levine et al.(1993)]{lev93} Levine, D., Beichmann, C., Fullmer, L., Helou, G.,
%Laughlin, G., Lord, S., Schmitz, M., and Surace, J. 1993, {\it IPAC User's Guide}, 5th Ed.

\bibitem[Looney, Mundy, \& Welch(2000)]{lo00} Looney, L., Mundy, L., and Welch, W. 2000, 
\apj, 529, 477

%\bibitem[Mardones et al.(1997)]{mar97} Mardones, D., Myers, P., Tafalla, M., Wilner, D.,
%Bachiller, R., and Garay, G. 1997, \apj, 489, 719

%\bibitem[Molinari et al.(1996)]{mol96} Molinari, S., Brand, J., Cesaroni, R.,
%and Palla, F. 1996, \aap, 308, 573

%\bibitem[Molinari et al.(1998a)]{mol98a} Molinari, S., Brand, J., Cesaroni, R.,
%Palla, F., and Palumbo, G. 1998, \aap, 336, 339

\bibitem[Molinari et al.(1998)]{mol98b} Molinari, S., Testi, L., Brand, J., Cesaroni, R.,
and Palla, F. 1998, \apj, 505, L39

\bibitem[Moriarty-Schieven et al.(1994)]{mor94} Moriarty-Schieven, G., Wannier, P., 
Keene, J., and Tamura, M. 1994, \apj, 436, 800

\bibitem[Myers \& Ladd(1993)]{mye93} Myers, P., and Ladd, E. 1993, \apj, 413, L47

\bibitem[Neckel \& Staude(1984)]{ne84} Neckel, T., \& Staude, H. 1984, \aap, 131, 200

\bibitem[O'Linger et al.(1999)]{oli99} O'Linger, J., Wolf-Chase, G., Barsony, M.,
and Ward-Thompson, D. 1999, \apj, 515, 696

%\bibitem[Persi, Palagi, and Felli(1994)]{per94} Persi, P., Palagi, F., and Felli, M.
%1994, \aap, 291, 577

\bibitem[Reipurth(2000)]{rei00} Reipurth, B. 2000, \aj, 120, 3177

\bibitem[Reipurth \& Clarke(2001)]{rei01} Reipurth, B., \& Clarke, C. 2001, \aj, 122, 432

\bibitem[Ressler \emph{et al.}(1994)]{ress94}Ressler, M. E., Werner, M. W.,
van Cleve, J., and Chou, H. A. 1994, Exp. Astr., 3, 277 

%\bibitem[Rice(1993)]{ric93} Rice, W. 1993, \aj, 105, 69

%\bibitem[Sandell et al.(1990)]{sa90} Sandell, G., Aspin, C., Duncan, W., Robson, E., and
%Dent, W. 1990, \aap, 232, 347

\bibitem[Saraceno et al.(1996)]{sar96} Saraceno, P., Andr\'{e}, P., Ceccarelli, C., Griffin,
M., \& Molinari, S. 1996, \aap, 309, 827

%\bibitem[Stahler, Shu, \& Taam(1980a)]{sta80a} Stahler, S., Shu, F., \& Taam, R. 1980a,
%\apj, 241, 637

\bibitem[Stahler, Shu, \& Taam(1980b)]{sta80b} Stahler, S., Shu, F., \& Taam, R. 1980,
\apj, 242, 226

%\bibitem[Saraceno et al.(1999)]{sar99} Saraceno, P., Benedettini, M., Di Giorgio, A.,
%Giannini, T., Nisini, B., Lorenzetti, D., Molinari, S., Pezzuto, S., Spinoglio, L., 
%Tommasi, E., Clegg, P., Correia, J., Griffin, M., Kaufman, M., Leeks, S., White, G.,
%Caux, E., Liseau, R., and Smith, H. 1999, in The Physics and Chemistry of the
%Interstellar Medium, ed. V. Ossenkopf et al., GCA-Verlag Herdecke, 279

%\bibitem[Stamatellos et al.(2005)]{sta05} Stamatellos, D., Whitworth, A.~P., Boyd, D.~F.~A., and Goodwin,
%S.~P. 2005, astro-ph 0505291 (to appear in \aap)

\bibitem[Tafalla et al.(2000)]{ta00} Tafalla, M., Myers, P., Mardones, D., and Bachiller,
R. 2000, \aap, 359, 967 

\bibitem[Terebey \& Padgett(1997)]{ter97} Terebey, S., and Padgett, D. 1997, in Herbig-Haro
Flows and the Birth of Low Mass Stars, ed. B.~Reipurth and C.~Bertout, IAU Conference Series 
182

\bibitem[Tsujimoto, Kobyashi, \& Tsuboi(2005)]{tsu05} Tsujimoto, M., Kobayashi, N., and Tsuboi, Y. 2005, 
astro-ph 0506628 (to appear in \aj)

\bibitem[Volgenau et al.(2002)]{vo02}  Volgnau, N., Mundy, L., Evans, N., and the Spitzer c2d
Legacy Project Team 2002, BAAS, 34, 1216

\bibitem[Wilking, Lada, \& Young(1989)]{wi89}  Wilking, B., Lada, C., and Young, E. 1989, \apj, 340, 823

\bibitem[Wolf-Chase, Barsony, \& O'Linger(2000)]{wo00a} Wolf-Chase, G.A., Barsony, M., and O'Linger, J. 
2000, \aj, 120, 1467

%\bibitem[Wolf-Chase et al.(2001)]{wo00b} Wolf-Chase, G.A., Moriarty-Schieven, G., 
%Fich, M., and Barsony, M. 2001, in prep.

\end{thebibliography}
\end{document}